\documentclass[11pt,a4paper,english,aps,preprint]{revtex4}
\usepackage[english]{babel}
\usepackage{graphicx,color}
\usepackage{latexsym}
\usepackage{amsmath}
\usepackage{amsfonts}
\usepackage{amssymb}
\usepackage[latin1]{inputenc}

\newcommand{\sech}{\rm sech}

\newcommand{\be}{\begin{equation}}
\newcommand{\ee}{\end{equation}}
\newcommand{\bea}{\begin{eqnarray}}
\newcommand{\eea}{\end{eqnarray}}

\begin{document}

\title{Multi-scalar tachyon potential on non-BPS domain walls}

\author{F. A. Brito}\email{fabrito@df.ufcg.edu.br}


\author{H. S. Jesuíno}\email{hjesuino@df.ufcg.edu.br}\affiliation{ Departamento de F\'\i
sica, Universidade Federal de Campina Grande, Caixa Postal 5008,
58051-970 Campina Grande, Para\'\i ba, Brazil}

\begin{abstract}
We have considered the multi-scalar and multi-tachyon fields living on a
3d domain wall embedded in a 5d dimensional Minkowski spacetime.
The effective action for such a domain wall can be found by integrating out the normal
modes as vibrating modes around the domain wall solution of a truncated 5d supergravity action.
The multi-scalar tachyon potential is good enough to modeling assisted inflation 
scenario with multi-tachyon fields. The tachyon condensation is also briefly addressed. 
\end{abstract}
\maketitle

\section{Introduction}

The superstring theory has in its non-perturbative spectrum objects such as BPS and non-BPS
D-branes. A D-brane-anti-D-brane system and the non-BPS D-branes have tachyons in their spectrum described by open strings \cite{1,2,3,4,5,6,7,8,9,Garousi:2007fk,Garousi:2008ge}.
The tachyon fields have shown to be involved in many scenarios with interesting physics such as inflationary cosmology and
particle physics. Being the D-brane dynamics governed by a gauge theory, a non-BPS D-brane theory in addition has 
the tachyon field that plays the role of giving mass to other fields, a phenomenon that indeed underlies the 
Higgs mechanism.

In string theory inflationary cosmology is hard to be implemented because it requires special compactifications to 
give rise to inflation \cite{10,11,12,13,14,15,16}. One can achieve this by considering a D3-brane-anti-D3-brane system in 
a warpped geometry (AdS space). The inflaton potential is related to the brane-anti-brane interaction. The potential found by 
embedding a brane-anti-brane system in AdS space is flatter than the potential of a brane-anti-brane system in a flat space \cite{11,12}. 
On the other hand, a system of N non-coincident non-BPS D3-branes can also implement inflation via multi-tachyon 
inflation \cite{Piao:2002vf}. As we shall show, using a similar setup to the latter case, we can also find sufficient flat potentials --- see also other recent alternatives using N D-branes \cite{Cai:2008if,Cai:2009hw,Zhang:2009gw,Li:2008fma}. 

In this paper we study a domain wall solution of the non-BPS sector of a five-dimensional supergravity theory. We assume
it can be found by suitable compactifications of type IIB supergravity. We look for tachyon modes in its
non-BPS sector. As we shall see we can find non-BPS domain wall solutions
where many tachyon fields can be found to live in their worldvolume.
We integrate out all the normal modes to find an effective action living in the domain wall worldvolume. This gives us
localized scalar fields. There is one massive, one massless, and a tower of tachyonic scalar modes for suitable choice 
of parameters \cite{zb,fb}. The effective action is similar to the action for the dynamics of N D3-branes at low energy with no charge. 
Each scalar mode in the action is then related to the location of a non-BPS D3-brane in the flat five-dimensional bulk. 

We show that at thin wall limit the action for the non-BPS domain wall world-volume is equivalent to the action
of $N$ decoupled tachyon fields. This may be related to tachyons living on the action for the world-volume of a
stack of $N$ non-coincident non-BPS D3-branes. The D3-branes
are far enough from each other such that tachyon modes from strings connecting distinct branes are absent. 

As we shall show, the tachyon potentials are flatter for larger tachyon masses. Since all the masses depend on the inverse of domain 
wall thickness, they become indeed large in the thin wall limit. Thus, in our scenario one can find 
sufficient flat potentials that can give rise to sufficient inflation.

The paper is organized as follows. In Sec.~\ref{II} we present the supergravity model and the domain wall solution and 
its fluctuation treating at first only three normal modes. In Sec.~\ref{III} we investigate the multi-scalar tachyon potential,
where we address the issues of tachyon condensation and Sen's conjecture. { One can show that the best result achieved is about $44.29\%$
of the expected answer. In Sec.~\ref{multiT} we extend the previous analysis to
five scalar modes, where new tachyon fields appear. The action for multi-tachyon fields and tachyon kink solutions are considered.
The cosmological implications are also considered. In Sec.~\ref{graviton} we address the issues of the gravitational field.
The complete effective four-dimensional action is found by considering the gravitational field that is naturally present 
in the bulk supergravity action.} In Sec.~\ref{IV} we make our final comments.

\section{The model}
\label{II}
Let us consider the bosonic part of a five-dimensional supergravity theory 
obtained
via compactification of a higher dimensional supergravity, given in the general form \cite{36,37,38,39,40}
\begin{equation}
\label{sugra}
e^{-1}{\cal L}_{sugra}=-\frac14M_*^3R_{(5)}+G_{AB}\partial_m\phi^A\partial^m\phi^B-\frac14 G^{AB}\frac{\partial W(\phi)}{\partial\phi^A}\frac{\partial W(\phi)}{\partial\phi^B}+\frac13\frac{1}{M_*^3}W(\phi)^2,
\end{equation}
where $e=|\det{g_{mn}}|^{1/2}$ and $G_{AB}$ is the metric on the scalar target space. $R_{(5)}$ is the 5d Ricci
scalar and $1/M_{*}$ is the five-dimensional Planck length. { For the sake of simplicity, below we consider the limit $M_*\gg1$
(with $M_*\ll M_{Pl}$) where five-dimensional gravity is not coupled to the scalar field. In this limit the non-BPS 3d domain wall is 
embedded in a five-dimensional Minkowski space. In Sec.~\ref{graviton}, we shall turn to the gravity field issues.}
In the present study we restrict the
scalar manifold to two fields only, i.e., $\phi_A = (\phi/\sqrt{2},\chi/\sqrt{2})$. 


Our supersymmetric Lagrangian (\ref{sugra}) can be now written in terms of 
two scalar fields, 
\begin{equation}
\label{A1.}  {\cal L} = \frac{1}{2}\partial_{m}\phi
\partial^{m}\phi + \frac{1}{2}\partial_{m}\chi
\partial^{m}\chi 
- V(\phi,\chi),
\end{equation}
where $V(\phi,\chi)=\frac{1}{2}\left(\frac{\partial W}{\partial\phi}\right)^2+\frac{1}{2}\left(\frac{\partial W}{\partial\chi}\right)^2$, being $W$ the superpotential. The theory is truncated up to two scalar fields. This shows to be enough to give rise to
a domain wall solution that can localize normal modes. The effective action for such modes is similar to the action of fields on a $D3$-brane worldvolume.

A simple choice of the superpotential is given by \cite{bnrt,morris,bb97,edels,shv,bbb,33}
\begin{equation}
\label{A2.}  W(\phi,\chi) = \lambda \left(\frac{\phi^{3}}{3} -
a^{2}\phi\right) + \mu \phi \chi^{2},
\end{equation}
such that the scalar potential developing a $Z_2\times Z_2$ symmetry is
\begin{equation}
\label{A3.}  V(\phi,\chi) = \frac{1}{2} \lambda^{2}(\phi^{2} -
a^{2})^{2}+ (2 \mu^{2}+ \lambda \mu) \phi^{2}\chi^{2} - \lambda \mu
a^{2}\chi^{2} + \frac{1}{2}\mu^{2}\chi^{4}.
\end{equation}
We shall determine later the effective action living on a non-BPS domain wall,
where the conventional particles 
are modes of the bulk scalar fields.

The scalar potential (\ref{A3.}) has the global minima
$(\phi = \pm a, \chi = 0)$ and $(\phi = 0, \chi = \pm a
\sqrt{\lambda/\mu})$. Each two vacua are connected by topological defects. These connected vacua comprise 
different topological sectors that have their own energy given by the Bogomol'nyi energy $E_B=|\Delta W|$
\begin{eqnarray}
\label{Aa.} (\phi = \pm a , \chi = 0), \quad\quad\quad  E_{B} =
\frac{4}{3}|\lambda|a^{3},
\end{eqnarray}
\begin{eqnarray}
\label{Ab.} (\phi = 0 , \chi = \pm a\sqrt{\lambda/\mu}),
\quad\quad\quad E_{B} = 0.
\end{eqnarray}
Note that the sector (\ref{Ab.}) is of the non-BPS type \cite{bnrt,morris,bb97,edels,shv,bbb,33}, since
its Bogomol'nyi energy is zero. Indeed, as we shall discuss later, this solution has a finite energy 
that can be properly found from the energy-momentum tensor. 
As one can be shown, the domain wall solutions 
coming from the BPS sector can localize fields and then another domain wall with smaller dimensions. 
However, we shall focus on the non-BPS sector, since multi-tachyon fields can also naturally be found.

The equations of motion for the 
scalar fields $\phi$ and $\chi$ are
\begin{equation}
\label{A4.}   \frac{d^2 \phi}{d x^2} = 2 \lambda^{2}(\phi^{2} -
a^{2})\phi  + 2 \mu^2\left(2 +
\frac{\lambda}{\mu}\right)\phi\chi^{2},
\end{equation}
\begin{equation}
\label{A5.}   \frac{d^2 \chi}{d x^2} = 2 \lambda\mu(\phi^2 -
a^2)\chi + 4 \mu^{2}\phi^{2}\chi + 2 \mu^{2}\chi^{3}.
\end{equation}
In the non-BPS sector $(\phi = 0 , \chi \neq 0)$ the equations of motion turn to
\begin{equation}
\label{A6.} \frac{d^{2} \chi}{d x^{2}} = 2 \mu^{2}\left(\chi^{2} -
\frac{\lambda a^{2}}{\mu}\right)\chi.
\end{equation}
A solution to this differential equation is given by
\begin{equation}
\label{A7.}   \chi = - a \sqrt{\frac{\lambda}{\mu}}
\tanh(a\sqrt{\lambda\mu} x), \quad \quad \quad \phi = 0,
\end{equation}
which we regard as the profile of a non-BPS 3d domain wall.

Calculating the energy of this solution we find $E_{nBPS} =
\frac{4}{3}|\lambda| a^{3}\sqrt{\frac{\lambda}{\mu}}$. This solution is stable as long as its energy 
is less than or equal to the energy of the BPS sector (\ref{Aa.}), i.e,
$E_{nBPS}\leq \frac{4|\lambda|a^{3}}{3}$. This ensure the non-BPS domain wall does not decay into a pair of BPS
defects \cite{bbb}. We conclude that this
solution is stable for $\frac{\lambda}{\mu} \leq 1$ --- See Ref.~\cite{6} for a similar discussion of a stable non-BPS D-string
of type IIA compactified on a orbifold. Thus,
for this solution decay into other settings (e.g., walls inside wall \cite{morris,bb97,edels,33}) is
necessary that $\frac{\lambda}{\mu} > 1$. The latter is the regime we are interested in and to which we shall turn our attention. This is
the regime where tachyon modes take place.

We consider the fluctuations around a general solution $(\overline{\phi}$, $\overline{\chi})$ in the form
\begin{equation}
\label{A8.}   \phi \rightarrow \overline{\phi} + \eta \quad \quad
\text{e}\quad \chi \rightarrow \overline{\chi},
\end{equation}
where $\eta$ describes the fluctuations of the field $\phi$ and reads 
\begin{equation}
\label{A9.}  \eta (t,x,y,z,w) = \sum_{n} \xi_{n}(t,y,z,w) \varphi_{n}(x).
\end{equation}

Let us now expand the action around the
solution (\ref{A7.}). Consider the transformation (\ref{A8.}) into the action 
as $S(\chi,\phi) \rightarrow S(\overline{\chi},\overline{\phi}
+ \eta)$ for the bosonic sector
\begin{equation}
\label{A10.}   S = \int d^4ydx{\cal L},
\end{equation}
and expand around the solution to obtain
\begin{eqnarray}
\label{A12.}   S &=& \int d^4ydx \left[
-\frac{1}{2}\left(\frac{d\overline{\chi}}{dx}\right)^2
  - V(\overline{\phi},\overline{\chi}) - \frac{1}{2}\partial_{\sigma}\eta\partial^{\sigma}\eta \right.
\nonumber\\
  &-& \left. \frac{1}{2}\eta\left(-\frac{\partial^2\eta}{\partial x^2} + \frac{\partial^2\overline{V}}{\partial\phi^2}\eta\right)
 - \frac{1}{6}\frac{\partial^{3}\overline{V}}{\partial\phi^3}\eta^3 - \frac{1}{24}\frac{\partial^{4}\overline{V}}{\partial\phi^4}\eta^4\right].
\end{eqnarray}
The first two terms of the expansion are responsible for the energy of non-BPS solution or simply the domain wall tension
\begin{eqnarray}
\label{A35.} T = \int_{-\infty} ^{+ \infty}dx \left
(\frac{1}{2}\left(\frac{d\overline{\chi}}{dx}\right)^2
 + V(\overline{\phi},\overline{\chi})\right)\equiv E_{nBPS}.
\end{eqnarray}
The fluctuations of the field $\phi$ is governed by the quadratic
$\eta$ terms of (\ref{A12.}). They provide a Schroedinger-like
equation for the fluctuations $\eta$ given as
\begin{equation}
\label{A13.} -\frac{\partial^2\eta}{\partial x^2} +
\frac{\partial^2\overline{V}}{\partial\phi^2}\eta = M_{n}^2 \eta.
\end{equation}
Now substituting Eq.~(\ref{A9.}) into Eq.~(\ref{A13.}),
developing the second derivative of potential and
defining $\tilde{x} = a\sqrt{\lambda \mu} x$, we obtain
\begin{equation}
\label{A15.} - \frac{d^{2}\varphi_{n}(\tilde{x})}{d
\tilde{x}^{2}} + \left[4 - \left(4 + 2
\frac{\lambda}{\mu}\right){\rm sech}^{2}(\tilde{x}) \right]
\varphi_{n}(\tilde{x}) = \frac{M_{n}^{2}}{a^{2} \lambda \mu}
\varphi_{n}(\tilde{x}).
\end{equation}
The equation (\ref{A15.}) is a solvable Schroedinger problem with a modified
Pöschl-Teller potential, whose eigenvalues are
\begin{equation}
\label{A16.} E_{n} = f - \left[\sqrt{g + \frac{1}{4}} - \left( n +
\frac{1}{2}\right)\right]^{2},
\end{equation}
with
\begin{equation}
\label{A17.} n = 0, 1, 2, 3, ... < \sqrt{g + \frac{1}{4}} -
\frac{1}{2},
\end{equation}
where $E_{n} = \frac{M_{n}^{2}}{a^{2} \lambda \mu}$, $f=
4$, $g = \left(4 + \frac{2 \lambda}{\mu}\right)$ and $n$ is the number
of bound states.

Assuming $\frac{\lambda}{\mu} >  1$ and using Eq.~(\ref{A17.}) we can determine the number of states present in our system
for each interval of $\frac{\lambda}{\mu}$. We present below the number
of states for some intervals of $\frac{\lambda}{\mu}$: 
\begin{eqnarray}
&&1 <
\frac{\lambda}{\mu} \leq 4, \qquad n = 0, 1, 2, 
\\ 
&&4 < \frac{\lambda}{\mu} \leq 8, \qquad
n = 0, 1, 2, 3, 
\\
&&8 < \frac{\lambda}{\mu} \leq 13, \qquad n = 0, 1, 2, 3, 4,
\end{eqnarray}
and so on. Note that the smallest number of bound states is three. Furthermore, for $2\lambda/\mu$ sufficiently large the modes are 
predominantly tachyonic. This is the case where we have a large number of tachyon fields $n\lesssim\sqrt{2\lambda/\mu}$ whose heavier 
tachyon has $\rm{mass}^2=-2/\Delta^2$, being $\Delta\sim1/\lambda a$ the domain wall thickness. This is similar 
to what happens in a stack of $N$ non-coincident parallel non-BPS $D3$-branes. We shall be back to this point later.

We can write the equation (\ref{A15.}) in the form
\begin{equation}
\label{A18.} -\frac{d^{2}\varphi_{n}(\tilde{x})}{d
\tilde{x}^{2}} + [\ell^{2} - \ell(\ell +1){\,\rm sech}^{2}
(\tilde{x})]\varphi_{n}(\tilde{x}) =
\omega_{n}^{2}\varphi_{n}(\tilde{x}),
\end{equation}
where we have used the following relations
\begin{equation}
\label{A19.} \frac{\lambda}{\mu} = \frac{\ell}{2}(\ell + 1) - 2,
\end{equation}
\begin{equation}
\label{A20.} \omega^{2}_{n} = \frac{M^{2}_{n}}{a^{2}\lambda \mu} +
(\ell^{2} - 4).
\end{equation}
The Schroedinger problem (\ref{A18.})
can be obtained by following the same method of Ref.~\cite{zb}. The eigenfunctions are given in terms of associated Legendre polynomials as $\ell$ is an integer. 

By following Eq.~(\ref{A19.}) we note that the condition $\lambda/\mu>1$ is satisfied only for $\ell\geq3$. Thus for $\ell = 3$ (i.e. $\frac{\lambda}{\mu} = 4$) we have the equation
\begin{equation}
\label{A21.} -\frac{d^2\varphi_{n}(\tilde{x})}{d
\tilde{x}^2} + \left[9 - 12{\,\sech^2}{(\tilde{x}})
\right]\varphi_{n}(\tilde{x}) = \left(\frac{4 M_{n}^2}{\lambda^2
a^2} + 5\right)\varphi_{n}(\tilde{x}).
\end{equation}
Solving this equation one can find the eigenfunctions and eigenvalues given
by
\begin{eqnarray}
\label{A22.} \varphi_{0}(x) &=& \sqrt{\frac{15a\lambda}{32}}{\rm
sech}^{3}\left(\frac{a\lambda}{2}x\right),\qquad\qquad\qquad\qquad\qquad M_{0}^2=
\frac{-5a^2\lambda^2}{4},
\\
\label{A23.} \varphi_{1}(x) &=&
\sqrt{\frac{15a\lambda}{8}}\tanh\left(\frac{a\lambda}{2}x\right){\,\rm
sech}^2 \left(\frac{a\lambda}{2}x\right),\qquad\quad\:\:\:\:\: M_{1}^2= 0,
 \\
 \label{A24.} \varphi_{2}(x) &=& \sqrt{\frac{3a\lambda}{32}}\left[5 {\,\rm sech}^3\left(\frac{a\lambda}{2}x\right) -
 4 {\,\rm sech}\left(\frac{a\lambda}{2}x\right)\right],\quad \:\:\:\:M_{2}^2= \frac{3a^2\lambda^2}{4},
\end{eqnarray}
being all the functions now displayed in their original variable $x$.
\begin{figure}[h]
  \begin{center}
        \includegraphics[scale=0.30]{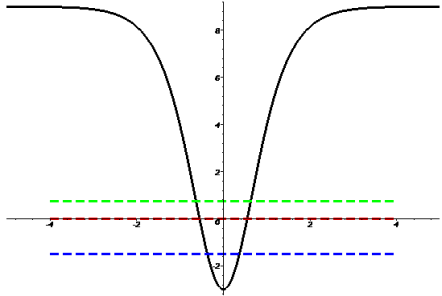}
  \end{center}
  \caption{The modified P\"{o}schl-Teller potential admitting three bound states.}
    \label{poten_poschl_telles}
\end{figure}

The potential in Eq.~(\ref{A21.}) admits three bound states --- see Fig.~\ref{poten_poschl_telles} --- such that
we can write the fluctuations $\eta(t, x, y,z)$ as 
\begin{eqnarray}
\label{A25.} \eta (t,x,y,z,w) = \xi_{0}(t,y,z,w)\varphi_{0}(x) +
\xi_{1}(t,y,z,w)\varphi_{1}(x) + \xi_{2}(t,y,z,w)\varphi_{2}(x).
\end{eqnarray}




Substituting Eqs.~(\ref{A22.})-(\ref{A25.}) in Eq.~(\ref{A12.}) and integrating out in $x$, we
obtain the effective action on the 3d $\chi$-domain wall, given by
\begin{eqnarray}
\label{A36.}   S_{eff} &=& \int d^4y \left[ - T
-\frac{1}{2}\sum_{n=0}^{2}\partial_{\sigma}\xi_{n}\partial^{\sigma}\xi_{n}
- V_{eff}(\xi_{n}) 
\right],
\end{eqnarray}
where $ T\equiv E_{nBPS}=
\frac{4}{3}|\lambda| a^{3}\sqrt{\frac{\lambda}{\mu}}$.
This is a theory of three real scalar fields living on a four-dimensional domain wall world-volume
that we get from a theory living in five-dimensions as we integrate out in the extra spatial dimension. 

\section{The multi-scalar tachyon potential}
\label{III}

As we mentioned above we should integrate out all the modes over the
extra spatial coordinate $x$ into the action (\ref{A12.}) in order to find
the effective action of the living modes on the world-volume of the
domain wall (\ref{A7.}). The multi-scalar tachyon potential living on the
world-volume reads
\begin{equation}
\label{A37.} V(\xi_{n}) = \int dx \left[-
\frac{1}{2}\eta\left(-\frac{\partial^2\eta}{\partial x^2} +
\frac{\partial^2\overline{V}}{\partial\phi^2}
 \eta\right) - \frac{1}{6}\frac{\partial^{3}\overline{V}}{\partial\phi^{3}}\eta^{3}
 - \frac{1}{24}\frac{\partial^{4}\overline{V}}{\partial\phi^4}\eta^4
 \right].
\end{equation}
Now substituting Eqs.~(\ref{A22.})-(\ref{A25.}) into Eq.(\ref{A37.}) and integrating out in $x$, 
we obtain the effective multi-scalar tachyon potential
\begin{eqnarray}
\label{A38.} V_{eff}(\xi_{0},\xi_{1},\xi_{2}) &=& -\frac{5 a^2
\lambda^2}{8}\xi_{0}^{2} + \frac{25}{154}a \lambda^3\xi_{0}^{4} +
\frac{3 a^2 \lambda^2}{8}\xi_{2}^{2} + \frac{9}{154}a
\lambda^3\xi_{2}^{4}+ \frac{15}{154}a \lambda^3\xi_{1}^{4}
\nonumber \\
&-& \frac{6\sqrt{5}}{77}a \lambda^3\xi_{0}\xi_{1}^2\xi_{2} +
\frac{30}{77}a \lambda^3\xi_{0}^{2}\xi_{1}^{2} + \frac{12}{77}a
\lambda^3\xi_{0}^{2}\xi_{2}^{2} + \frac{18}{77}a
\lambda^3\xi_{1}^{2}\xi_{2}^{2}
\nonumber \\
&-& \frac{4\sqrt{5}}{385}a \lambda^3\xi_{0}\xi_{2}^{3} +
\frac{6\sqrt{5}}{77}a \lambda^3\xi_{0}^{3}\xi_{2}.
\end{eqnarray}
The coupling among the fields are controlled by the domain wall tension $T\sim\lambda a^3$ and the squared mass of 
the fields can be given in terms of the domain wall thickness $\Delta\sim 1/\lambda a$. The tachyon in superstring theory
coming from non-BPS D-branes has ${\rm mass}^2=-1/2\alpha'$. By identifying this tachyon with ours given in the potential (\ref{A38.})
we conclude that $\sqrt{\alpha'}\sim \Delta$. One can use this thickness as a parameter to 
control the coupling among the fields. For instance, for sufficiently thin domain wall ($\lambda a\to\infty$, $\lambda a^3\to$ fixed),
the quadratic terms dominate over the coupling and self-coupling terms, such that the potential for $N$ modes is approximately given by
the sum of $N$ independent potentials
\begin{equation}
\label{decoup_pots}
V_{eff}(\xi_{0},\xi_{1},...,\xi_{N})=V_0(\xi_{0})+V_1(\xi_{1})+...+V_N(\xi_{N}),
\end{equation}  
with $V_n(\xi_n)=\frac12M^2_{n}\xi_n^2$, being $M_n^2$ given by the Schroedinger problem (\ref{A18.}).
On the other hand, for sufficiently thick domain walls, the coupling among the fields take place and the squared masses 
change. Let us assume some values for the parameters of our theory to determine the masses of real scalar
fields ($\xi_{0}$, $\xi_{1}$ e $\xi_{2}$) as they interact. For $\lambda=4$ and $a=0.15$ we find the following masses $m_{\xi_{1}}=0.419666$,
$m_{\xi_{2}}=0.639094$ and $m_{\xi_{0}}=1.09152$ by diagonalizing the matrix of squared masses. Note the ``tachyon condensation''
via the tachyon field $\xi_0$ since it now has a positive mass. 

As a first approximation, we consider the multi-scalar tachyon
potential at level zero where only the tachyon field $\xi_{0}$ is
present
\begin{equation}
\label{A39.} V_{0}(\xi_{0}) = -\frac{5 a^2 \lambda^2}{8}\xi_{0}^{2}
+ \frac{25}{154}a \lambda^3\xi_{0}^{4}.
\end{equation}
In analogy with Sen's conjecture in string theory \cite{5,6} we can test the validity of the condition $T + V(\xi^{*}) = 0$,
at $\xi=\xi^{*}$, a regime where the non-BPS Dp-brane is indistinguishable from the vacuum with no D-brane.

Let us investigate how much $V(\xi^{*})$ approaches the tension $T$. At level zero the result is independent of the parameters $\lambda, a$.
The nontrivial critical points are $\xi_{0}^{*}= \pm \sqrt{\frac{77 a}{40\lambda}}$ in which
the potential assumes the absolute value $|V_{0}(\xi^{*})| =
\frac{77a^{3}\lambda}{128}$. The domain wall tension is $T = \frac{8
a^{3}\lambda}{3}$, thus
\begin{equation}
\label{A40.} \frac{|V_{0}(\xi^{*})|}{T} = 0.22559,
\end{equation}
which corresponds to about 22.56\% of the expected result. In the following we consider the next level.

Adding the massless scalar field $\xi_{1}$ to the tachyon potential
(\ref{A39.}), we find that $|V_{0}(\xi_{0}^{*},\xi_{1}^{*})|/T$ does not change.
Now including the massive field $\xi_{2}$ we obtain the multi-scalar tachyon
potential
\begin{eqnarray}
\label{A41.} V(\xi_{0},\xi_{2}) &=&  -\frac{5 a^2
\lambda^2}{8}\xi_{0}^{2} + \frac{25}{154}a \lambda^3\xi_{0}^{4} +
\frac{3 a^2 \lambda^2}{8}\xi_{2}^{2} + \frac{9}{154}a
\lambda^3\xi_{2}^{4} + \frac{12}{77}a
\lambda^3\xi_{0}^{2}\xi_{2}^{2}
\nonumber \\
&-& \frac{4\sqrt{5}}{385}a \lambda^3\xi_{0}\xi_{2}^{3} +
\frac{6\sqrt{5}}{77}a \lambda^3\xi_{0}^{3}\xi_{2}.
\end{eqnarray}
This potential has the nontrivial critical points $\xi_{0}^{*}
\simeq \pm 4.3257$, $\xi_{2}^{*} \simeq \mp 1.13075$, where
$|V(\xi_{0}^{*},\xi_{2}^{*})| \simeq 22.43078$ --- we have assumed $\lambda = 0.5$ and $a = 4$. 
Thus, at this level we find
\begin{equation}
\label{A42.} \frac{|V(\xi_{0}^{*},\xi_{2}^{*})|}{T} = 0.26286,
\end{equation}
that corresponds to about 26.29\% of the expected result. This make a little improvement of the result obtained at level zero.

{ \subsection{The fluctuations of the scalar field $\chi$}
\label{fluct_chi}

All we have done until now can be easily repeated for the fluctuations of the field $\chi$. For the non-BPS domain wall, the Schroedinger-like equation for the fluctuations of the field $\chi$, $\zeta(x,y)=\sum_n\Psi_n(x)\tau_n(x)$, decouples from the fluctuations of the field $\phi$, $\eta(x,y)$,  that we have previously considered, and reads
\bea
-\frac{d^2\Psi_n(\tilde{x})}{d\tilde{x}^2}+[4-6\,{\sech^2}{(\tilde{x})}]\Psi_n(\tilde{x})=\frac{4M_n^2}{a^2\lambda^2}\Psi_n(\tilde{x}).
\eea
In this eigenvalue problem we find only two bound states with squared masses $M_0^2=0$ and $M_1^2=\frac{3\lambda^2 a^2}{4}$ with eigenfunctions given by
\bea
\Psi_0(x)=\sqrt{\frac{3\lambda a}{8}}\,{\sech}^2\left(\frac{a\lambda}{2}\,x\right), \qquad \Psi_1(x)=\sqrt{\frac{3\lambda a}{4}}\,{\tanh}\left(\frac{a\lambda}{2}\,x\right)
\,{\sech}\left(\frac{a\lambda}{2}\,x\right).
\eea
The effective potential now has two more interacting scalar modes
\bea
\label{veff}
V_{eff}(\xi_0,\xi_1,\xi_2,\tau_0,\tau_1)=V_{eff}(\xi_0,\xi_1,\xi_2)+\Delta V_{eff},
\eea
where the first part is precisely the potential in Eq.~(\ref{A38.}) and $\Delta V_{eff}$ comprise the new modes $\tau_0$ and $\tau_1$
\bea
\Delta V_{eff}&\!=\!&{\frac {3}{28}}\,{\lambda}^{3}a{{\tau_0}}^{2}{{\xi_0}}^{2}+{\frac 
{3}{140}}\,{\lambda}^{3}a{{\tau_0}}^{2}{{\xi_2}}^{2}+{\frac {9}{
560}}\,{\lambda}^{3}a{{\tau_0}}^{2}{{\tau_1}}^{2}+{\frac {3}{56}}
\,{\lambda}^{3}a{{\tau_0}}^{2}{{\xi_1}}^{2}+{\frac {3}{80}}\,{
\lambda}^{3}a{{\tau_1}}^{2}{{\xi_2}}^{2}\nonumber\\
&+&{\frac {3}{112}}\,{\lambda}^{3}a
{{\tau_1}}^{2}{{\xi_0}}^{2}+{\frac {9}{2240}}\,{\lambda}^{3}a{{
\tau_1}}^{4}+\frac{3{a}^{2}{\lambda}^{2}}{8}\,{{\tau_1}}^{2}+{\frac {9}{
1120}}\,{\lambda}^{3}a{{\tau_0}}^{4}+{\frac {3\,\sqrt {2}}{28}}\,{\lambda}^{3}a
{\tau_0}{\tau_1}\,{\xi_0}\,{\xi_1}\nonumber\\
&+&{\frac {135\,\pi}{
2048}}\sqrt{{3\lambda}^{5}{a}^{3}}{\tau_1}{{\xi_1}}^{2}
 -{\frac {63\,\pi}{4096}}\sqrt{{30\lambda}^{5}{a}^{3}}{\tau_1}{\tau_0}\,{\xi_1}
{\xi_2} -{\frac {63\,\pi}{4096}}\sqrt{{15\lambda}^{5}{a}^{3}}{\tau_1}{\xi_0}
\xi_2 +{\frac {3}{56}}\,{\lambda
}^{3}a{{\tau_1}}^{2}{{\xi_1}}^{2}\nonumber\\
&+&{\frac {9\,\pi}{512}}\sqrt{{3\lambda}^{5}{a}^{3}}{{\tau_0}}^{2}{
\tau_1} +{\frac {3\,\pi}{256}}\sqrt{{3\lambda}^{5}{a}^{3}}{{
\tau_1}}^{3} +{\frac {3\sqrt {5}}{140}}\,{\lambda}^{3}a{{\tau_0
}}^{2}{\xi_0}{\xi_2}+{\frac {225\,\pi}{4096}}\sqrt{{6\lambda}^{5}{a}^{3}}{\tau_0}{\xi_0}\,{\xi_1} \nonumber\\
&+&{
\frac {549\,\pi}{8192}}\sqrt{{3\lambda}^{5}{a}^{3}}{\tau_1}{{
\xi_2}}^{2} -{\frac {3\,\sqrt {10}}{140}}\,{\lambda}^{3}a{\tau_0}\,{
\tau_1}\,{\xi_1}{\xi_2}-{\frac {3\sqrt {5}}{280}}{
\lambda}^{3}a{{\tau_1}}^{2}{\xi_0}{\xi_2}+{\frac {225\,\pi
}{8192}}\sqrt{{3\lambda}^{5}{a}^{3}}{\tau_1}{{\xi_0}}^{2}. 
\eea
Thus, the potential (\ref{veff}) has the nontrivial critical point $ 
\xi^*_0 \simeq 5.6621, \xi^*_1 = 0, \xi^*_2 \simeq -3.6864, \tau^*_0 = 0, \tau^*_1 \simeq -5.2661$
such that
\begin{equation}
\label{A42.20} \frac{|V(\xi_{0}^{*},\xi_{1}^{*},\xi_{2}^{*},\tau_{0}^{*},\tau_{1}^{*})|}{T} = 0.44293.
\end{equation}
This corresponds to about 44.29\% of the expected result --- again, we have assumed $\lambda = 0.5$ and $a = 4$. This make a good improvement of the result obtained at level zero, but it is the
best result one can achieve in this model.
}

In the sense of achieving a better result we could attempt to take into account more fields from the ``$\phi$ - sector''. However, as we shall see, the next modes developed in the present model
are tachyon fields. Thus our non-BPS domain wall is pretty much like related to $N$ non-BPS D-branes. In the next section these
additional tachyon fields are considered.  
{ 
We shall mainly focus on the tachyon modes localized on the non-BPS domain wall.}

\section{Multi-tachyon fields}
\label{multiT}



\begin{figure}[h]
     \includegraphics[scale=0.30]{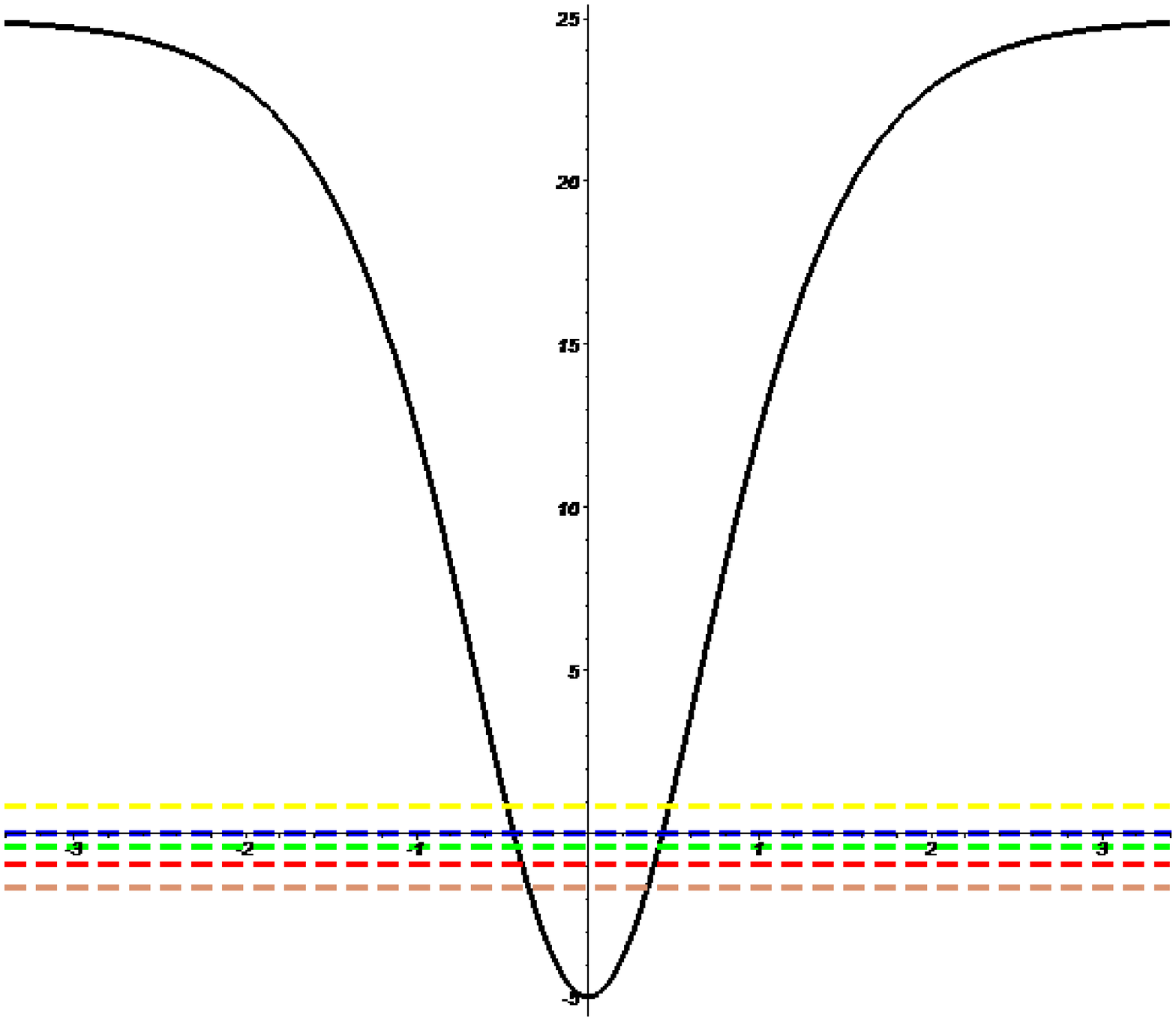}
        \caption{The modified P\"{o}schl-Teller potential admitting five bound states.}
        \label{potencial_2}
\end{figure}

As we have earlier discussed we can choose the number of modes $\xi_{n}$ by
properly setting a value for $\frac{\lambda}{\mu}>1$. Furthermore, analyzing the Schroedinger 
problem for our model, we observe that as we increase $\frac{\lambda}{\mu}$ the potential deepens, 
such that we have a larger number of modes. These new modes will be necessarily tachyon modes, as
depicted in Fig.~\ref{potencial_2}. 

As an example, we now consider a theory with three tachyon modes. To obtain 
a potential that supports them we admit that
$\frac{\lambda}{\mu} =  13$ in equation (\ref{A15.}). This will make to appear five 
scalar modes.
Thus for $\frac{\lambda}{\mu} = 13$ we have the Schroedinger problem
\begin{equation}
\label{A21.2} -\frac{d^2\varphi_{n}(\tilde{x})}{d
\tilde{x}^2} + \left[25 - 30{\,\sech^2}{(\tilde{x}})
\right]\varphi_{n}(\tilde{x}) = \left(\frac{13 M_{n}^2}{\lambda^2
a^2}+21\right)\varphi_{n}(\tilde{x}).
\end{equation}
Solving this equation one can find the eigenfunctions given
by
\begin{eqnarray}
\label{five}
&&\varphi_{0}(x) = \sqrt{\frac{315 a \lambda \sqrt{13}}{3328}}{\,\rm
sech}^{5}\left(\frac{a \lambda \sqrt{13}}{13}x\right), 
\nonumber\\
&&\varphi_{1}(x) = \sqrt{\frac{315 a \lambda \sqrt{13}}{416}}{\,\rm
sech}^{4}\left(\frac{a \lambda
\sqrt{13}}{13}x\right)\tanh\left(\frac{a \lambda
\sqrt{13}}{13}x\right), 
\nonumber\\
&&\varphi_{2}(x) = \sqrt{\frac{105 a \lambda
\sqrt{13}}{3328}}\left[9{\,\rm sech}^{5}\left(\frac{a \lambda
\sqrt{13}}{13}x\right) - 8{\,\rm sech}^{3}\left(\frac{a \lambda
\sqrt{13}}{13}x\right)\right], 
\\
&&\varphi_{3}(x) = \sqrt{\frac{35 a \lambda
\sqrt{13}}{39936}}\left[72{\,\rm sech}^{4}\left(\frac{a \lambda
\sqrt{13}}{13}x\right)\tanh\left(\frac{a \lambda
\sqrt{13}}{13}x\right)\right.
\nonumber\\
&& \left.\qquad\qquad\qquad\qquad\qquad\qquad -48{\,\rm sech}^{2}\left(\frac{a \lambda
\sqrt{13}}{13}x\right)\tanh\left(\frac{a \lambda
\sqrt{13}}{13}x\right)\right], 
\nonumber\\
&&\varphi_{4}(x) = \sqrt{\frac{5 a \lambda
\sqrt{13}}{19968}}\left[48{\,\rm sech}\left(\frac{a \lambda
\sqrt{13}}{13}x\right) - 168{\,\rm sech}^{3}\left(\frac{a \lambda
\sqrt{13}}{13}x\right) + 126{\,\rm sech}^{5}\left(\frac{a \lambda
\sqrt{13}}{13}x\right)\right]. 
\nonumber
\end{eqnarray}
All the functions are now displayed in their original variable $x$.
The eigenvalues are $M^{2}_{0} = -\frac{21 }{13}a^{2}\lambda^{2}$, $M^{2}_{1} = -\frac{12}{13}a^{2}\lambda^{2}$, $M^{2}_{2} = -\frac{5}{13}a^{2}\lambda^{2}$, $M^{2}_{3} = 0$ and $M^{2}_{4} = \frac{3}{13}a^{2}\lambda^{2}$, respectively.

The potential in Eq.~(\ref{A21.2}) admits five bound states --- see Fig.~\ref{potencial_2} --- such that
we can write the fluctuations $\eta(t, x, y,z,w)$ as 
\begin{eqnarray}
\label{A25.2} \eta (t,x,y,z,w)\! &=& \!\xi_{0}(t,y,z,w)\varphi_{0}(x)\!+\!
\xi_{1}(t,y,z,w)\varphi_{1}(x)\! + \!\xi_{2}(t,y,z,w)\varphi_{2}(x)\!\nonumber\\
&+&\!\xi_{3}(t,y,z,w)\varphi_{3}(x)\!+\!\xi_{4}(t,y,z,w)\varphi_{4}(x).\:\:\:\:
\end{eqnarray}




As in the previous example, substituting Eqs.~(\ref{five})-(\ref{A25.2}) into Eq.~(\ref{A12.}) and integrating out in $x$, we
obtain the effective action on the 3d $\chi$-domain wall, given by
\begin{eqnarray}
\label{A36.}   S_{eff} &=& \int d^4y \left[ - T
-\frac{1}{2}\sum_{n=0}^{4}\partial_{\sigma}\xi_{n}\partial^{\sigma}\xi_{n}
- V_{eff}(\xi_{n}) 
\right].
\end{eqnarray}
We find an action describing the dynamics of five scalar fields with a multi-scalar tachyon potential
 $V_{eff}(\xi_{0},\xi_{1},\xi_{2},\xi_{3},\xi_{4})$. Due to the existence of many tachyons, this action
should be related to many non-BPS D-branes.


As we have anticipated, for sufficiently large $2\lambda/\mu$, the modes are predominantly tachyon modes. Furthermore, in the thin domain wall limit, i.e., $\lambda a\to\infty$, these fields are very weakly coupled such that the multi-tachyon potential can be written as in Eq.~(\ref{decoup_pots}).
Thus, we can write the effective action (\ref{A36.}) for $N$ tachyon fields in the form
\begin{eqnarray}
\label{multi-tach}   S_{eff} &=& \sum_{n=0}^{N}\int d^4y\left[ - T_{(3)n}
-\frac{1}{2}\partial_{\sigma}\xi_{n}\partial^{\sigma}\xi_{n}
- V_{n}(\xi_{n})
\right],
\end{eqnarray}
where we have also identified the domain wall tension $T_{(3)}=T_{(3)1}+T_{(3)2}+...+T_{(3)N}$. To make contact with D-branes we should add higher derivatives and write this action in the DBI-like form
\begin{eqnarray}
\label{DBI}
S=-\sum_{n=0}^N\int{d^4y V_n(T_n)\sqrt{1+\partial_\mu T_n\partial^\mu T_n}}.
\end{eqnarray}
This can be achieved by redefining the fields $\xi_n$ and the potentials $V_n(\xi_n)$ as follows
\begin{eqnarray}
\label{redef_pot}
V_n(\xi_n)+T_{(3)n}=\left[\frac{\partial\xi_n(y)}{\partial T_n}\right]^2=V_n(T_n),\\
\label{redef_field}
\partial_\sigma\xi_n(T_n(y))=\frac{\partial\xi_n(y)}{\partial T_n}\partial_\sigma T_n(\xi_n(y)).
\end{eqnarray}
Now substituting into Eq.~(\ref{multi-tach}) we find
\begin{eqnarray}
\label{multi-tach2}S_{eff} &=& \sum_{n=0}^{N}\int d^4y \left[ - V_n(T_n)-\frac12V(T_n)\partial_\sigma T_n \partial^\sigma T_n
\right]
\end{eqnarray}
Note this is the action (\ref{DBI}) expanded up to quadratic first derivative terms (low energy limit). Thus, in order to take into account all the higher derivatives 
we should consider the action in the form (\ref{DBI}). As we have earlier discussed, in the thin domain wall approximation we have quadratic 
potentials $V_n(\xi_n)=-\frac12|M_n|^2\xi_n^2$. By following the definitions (\ref{redef_pot}) and (\ref{redef_field}), we find 
\begin{equation}V_n(T_n)=T_{(3)n}\left(1-\frac{\sin{(\sqrt{2}\,T_n|M_n|/{2})}^2}{|M_n|^2}\right),\end{equation} where $\xi_n(T_n)=\left(\sqrt{2\,T_{(3)n}}/|M_n|\right)\sin{\left(\sqrt{2}\,T_n|M_n|/{2}\right)}$, being $M_n^2$ the squared tachyon masses. These potentials are periodic with period $2(\sqrt{2}/|M_n|)\arcsin{(|M_n|)}$. Note that for heavier tachyons ($|M_n|\to\infty$) the tachyon potentials become 
flatter (i.e., $V_n(T_n)\simeq T_{(3)n}$ at small tachyon fields $T_n$) such that they can provide sufficient inflation. Indeed, in the thin wall limit,
all the tachyon masses are large since $|{\rm mass}^2|\sim\lambda^2a^2\to\infty$. For flat potentials $V_n(T_n)$ (i.e., almost constants) the DBI-like action (\ref{DBI})
returns us the solutions $T(y)\propto y$, such that we find the well-known kink solution \cite{6}
\begin{equation}
\xi_n(y)\sim \sin{\left(\sqrt{2}\,|M_n|y/{2}\right)}.
\end{equation}

\subsection{Tachyon kinks}

\begin{figure}[h]
  \begin{center}
        \includegraphics[scale=0.30]{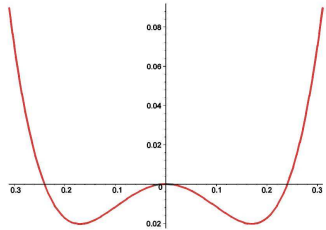}
  \end{center}
  \caption{The tachyon potential at level zero.}
    \label{poten_tachyon}
\end{figure}

We can also restrict ourselves to the study of a tachyon potential at level zero. 
Let us use the tachyon potential (\ref{A39.}) --- see Fig.~\ref{poten_tachyon}.
Note that this tachyon potential is invariant under a $Z_{2}$-symmetry.
It supports a tachyon kink interpolating between the two minima of the potential, $\xi_{0}^{*}= \pm
\sqrt{\frac{77 a}{40\lambda}}$. This kink is the profile of a BPS domain wall
with a dimension smaller than the dimension of the non-BPS 3d domain wall. This is analogous to the case of 
a BPS D$(p-1)$-brane with tension $T_{p-1}$ that lives on a non-BPS
D$p$-brane with tension $T_{p}$ \cite{5,6}. Thus, using the effective action living on the 3d domain wall (\ref{A36.}), 
we obtain the kink solution as the profile of a 2d domain wall with tension $T_{2}$, i.e.,
\begin{equation}
\label{A43.} \xi_{0} = \pm \xi_{0}^{*}\tanh(y), \quad\quad\quad\quad
T_{2} = \frac{4}{3}(\xi_{0}^{*})^{2}.
\end{equation}

\section{Turning on the gravitational field}
\label{graviton}
In the preliminaries sections, for the sake of simplicity, we adopted the flat limit of the five-dimensional space
and did not address the gravitational field that is naturally present in supergravity theories. Thus, sufficiently far
from the flat regime our four-dimensional effective action for the multi-tachyon fields we present above is 
incomplete. The complete effective four-dimensional action is found by considering the gravitational field that is naturally present 
in the bulk supergravity action and them integrate out the scalar modes and gravitational field.
This complete action is crucial to address cosmological issues in the four-dimension effective action for the multi-tachyon fields  \cite{Piao:2002vf,Cai:2008if,Cai:2009hw}. 

Let us now write the bosonic part of a five-dimensional supergravity, through the Lagrangian (\ref{sugra}), in the following form
\bea
\label{sugra2}
S=\int{d^4ydx\sqrt{|g^{(5)}|}\left(-\frac14 R_{(5)}+{\cal L}\right)}.
\eea
Here without any loss of generality we have set $M_*=1$.
The Lagrangian ${\cal L}$ for the scalar fields is
formally the same as in Eq.~(\ref{A1.}), but the potential is now given by
\bea
\label{sugra_pot}
V(\phi,\chi)=\frac12\left(\frac{\partial W}{\partial \phi}\right)^2+\frac12\left(\frac{\partial W}{\partial \chi}\right)^2-\frac13W^2.
\eea
The metric of the five-dimensional spacetime assumed to have a four-dimensional Poincare invariance along the domain wall is given 
in the form
\bea
\label{metric}
ds^2=e^{2A(x)}\eta_{\mu\nu}dx^\mu dx^\nu-dx^2, \qquad \mu,\nu=0,1,2,3.
\eea
Admitting that both $A$ and scalar fiels $\phi,\chi$ depend only on the fifth coordinate $x$, then the equation of motion for the scalar fields
and the Einstein's equations read \cite{36,37,Karch:2000ct,38,39,40,Bazeia:2006ef,Bazeia:2004yw,Bazeia:2004dh}
\bea
\label{eoms}
&&\phi''+4A'\phi'=\frac{\partial{V}}{{\partial\phi}}, \qquad \chi''+4A'\chi'=\frac{\partial{V}}{{\partial\chi}},\nonumber\\
&&A''=-\frac23\phi'^2-\frac23\chi'^2, \qquad A'^2=\frac16\phi'^2+\frac16\chi'^2-\frac13V(\phi,\chi),
\eea
where prime means derivatives with respect to $x$. 

In the non-BPS sector $\phi=0$ and $\chi\neq0$, the superpotential vanishes 
and so does the Bogomol'nyi energy as in our previous analysis. In this sector the Eqs.~(\ref{eoms}) reduce to 
\bea
\label{eom0}
&&\chi''+4A'\chi'=2\mu^2\left(\chi^2-\frac{\lambda a^2}{\mu}\right)\chi\nonumber\\
&&A''=-\frac23\chi'^2, \qquad A'^2=\frac16\chi'^2-\frac16\mu^2\left(\chi^2-\frac{\lambda a^2}{\mu}\right)^2.
\eea
The non-BPS domain wall solution we have considered above still satisfies the Eqs.~(\ref{eom0}) as long as we neglect
the term $4A'\phi'$ --- compare with Eq.~(\ref{A5.}). This is readily true in the thin wall limit we have earlier discussed. In this limit we can approach the kink solution to a step function $\chi\simeq\sqrt{\lambda/\mu}a$ sgn$(x)$, whose width $\Delta\simeq1\sqrt{\lambda\mu}a$ goes to zero. This allows us to make use of the ``identities'' \cite{38,Bazeia:2004yw}: $\chi'\simeq2\sqrt{\lambda/\mu}a\,\delta(x)$
and $\chi'^2\simeq2\sigma\delta(x)$, where $\sigma\equiv T_{(3)}$ is the domain wall tension. One should also add a negative cosmological constant to the potential, i.e., $V(0,\chi)\to V(0,\chi)-3/L^2$, that it will be justified shortly. We have that $A'\phi'=2\sqrt{\lambda/\mu}a\, A'(x)\delta(x)$ is zero everywhere provided that $A(x)$ satisfies the boundary condition $A'(0)=0$. Thus the Eqs.~(\ref{eom0}) now turn to 
\bea
\label{eom2}
A''=-\frac{2}{3}\sigma\delta(x), \qquad A'^2=\frac{1}{L^2}.
\eea
This gives us the solution $A(x)=-|x|/L$, where $L=3/\sigma$ is $AdS_5$ radius since asymptotically this solution describes
an $AdS_5$ space. 

Let us now verify our previous assumption concerning the cosmological constant on the bulk. Our non-BPS solution itself 
cannot produce such a constant. { However, this is expected to come from the fluctuations of the scalar fields $\phi$ and $\chi$. Recall that we have assumed such fluctuations $\delta\phi=\eta(x,y)\equiv\sum_n\xi_n(y)\psi_n(x)$ and $\delta\chi=\zeta(x,y)\equiv\sum_n\tau_n(y)\Psi_n(x)$.

The equations (\ref{eoms}) can give us an answer about the effect of such scalar fluctuations on the metric by considering the following: $A(x)\to A(x,y)=A(x)+\delta A(x,y)$, where $\delta A(x,y)$ will correspond to our first order correction on the metric. 

Now making a functional variation of $A''$ in Eqs.~(\ref{eoms}), we find
\bea
\label{deltaA}
(\delta A)''=-\frac43\phi'\delta\phi'-\frac43\chi'\delta\chi' \to  -\frac43\chi'\partial_x \zeta(x,y) \to -\frac83\sqrt{\frac{\lambda}{\mu}}
a\,\delta(x)\partial_x\zeta(x,y),
\eea 
where we have used $\delta A''=(\delta A)''$ and $\delta \chi'=(\delta \chi)'=\partial_x\zeta(x,y)$. Recall that for the non-BPS domain wall solution 
$\phi=0$ and $\phi'=0$, and in the thin wall limit $\chi'\simeq2\sqrt{\frac{\lambda}{\mu}}a\delta(x)$. One can easily integrate (\ref{deltaA}) to find 
\bea
\label{sol_deltaA}
\delta A(x,y)=-\frac43\sqrt{\frac{\lambda}{\mu}}a\,\zeta'(0,y)|x|=-k\,\zeta'(0,y)|x|.
\eea


The fluctuations of the gravitational field around the flat space can be now addressed by using the metric in the form \cite{Randall:1999ee}
\bea
\label{metric_fluct}
ds^2=e^{-2k|X|}[\eta_{\mu\nu}+\bar{h}_{\mu\nu}(y)]dx^\mu dx^\nu-dX^2, \qquad dX^2=\zeta'(0,y)^2dx^2,
\eea
where $\zeta'(0,y)=\partial_x\zeta(x,y)|_{x=0}$.
Note that $\zeta'(0,y)=\sum_{n}{a_n\tau_n(y)}$ can be viewed as a summation on ``moduli fields" $\tau_n(y)$ with $a_n=\Psi'_n(0)$. They are stabilized at their vacuum expectation value $\tau^*_n$ \cite{Goldberger:1999uk}. As we have shown, there is a configuration of vacuum as follows: $ 
\xi^*_0 \simeq 5.6621, \xi^*_1 = 0, \xi^*_2 \simeq -3.6864, \tau^*_0 = 0, \tau^*_1 \simeq -5.2661$. Finally from Eqs.~(\ref{eoms}), we also find $(\delta A)'^2=k^2a_1^2{\tau^*_1}^2\equiv1/L^2$, with $L\sim 1/\sigma$, that ensures asymptotically the existence of a 5d negative cosmological constant $\Lambda=-3/L^2$.}

As we have earlier discussed, the action (\ref{multi-tach}) comes from (\ref{sugra2}), in the 5d flat space limit, integrated out in the fifth dimension $x$. It can also be written as
\begin{eqnarray}
\label{multi-tach-2}   S_{eff} &=& \sum_{n=0}^{N}\int d^4ydx\,\delta(x)\sqrt{|g^{(5)}|}\left[ - T_{(3)n}
-\frac{1}{2}\partial_{\sigma}\xi_{n}\partial^{\sigma}\xi_{n}
- V_{n}(\xi_{n})
\right],\nonumber\\
&=& \sum_{n=0}^{N}\int d^4y\sqrt{|g^{(4)}|}\left[ - T_{(3)n}
-\frac{1}{2}\partial_{\sigma}\xi_{n}\partial^{\sigma}\xi_{n}
- V_{n}(\xi_{n})
\right],
\end{eqnarray} where $g^{(4)}_{\mu\nu}(y)\equiv g^{(5)}_{\mu\nu}(0,y)=\eta_{\mu\nu}$. In the flat space limit, the curvature term does not contribute. However, as we early observed, in order to take into account the effects of the  fluctuations of the scalar fields $\phi$ and $\chi$, 
one must work with the fluctuations of the metric too, as stated in (\ref{metric_fluct}). Thus, the effective action now reads
\begin{eqnarray}
\label{multi-tach-3}  S_{eff} &=&\frac12 M_{Pl}^2\int d^4y\sqrt{|\bar{g}^{(4)}|}\bar{R} +\sum_{n=0}^{N}\int d^4y\,\sqrt{|\bar{g}^{(4)}|}\left[ - T_{(3)n}
-\frac{1}{2}\partial_{\sigma}\xi_{n}\partial^{\sigma}\xi_{n}
- V_{n}(\xi_{n})
\right],
\end{eqnarray}
where $\bar{g}^{(4)}_{\mu\nu}\equiv g^{(5)}_{\mu\nu}(0,y)=\eta_{\mu\nu}+\bar{h}_{\mu\nu}(y)$ and the curvature term is now made out of the metric $\bar{g}^{(4)}_{\mu\nu}$ \cite{Randall:1999ee}. The extension of the multi-tachyon part into a DBI-like form as in (\ref{DBI}) is straightforward: 
\begin{eqnarray}
\label{DBI-2}
S_{eff}=\frac{1}{2\kappa^2}\int d^4y\sqrt{|\bar{g}^{(4)}|}\bar{R}-\sum_{n=0}^N\int{d^4y\,\sqrt{|\bar{g}^{(4)}|} V_n(T_n)\sqrt{1+\partial_\mu T_n\partial^\mu T_n}},
\end{eqnarray}
where $\kappa^2=1/M_{Pl}^2$. This action is the start point of the multi-tachyon cosmology settings.
\section{Conclusions}
\label{IV}

We have shown that the non-BPS sector of a five-dimensional supergravity theory can give us an effective theory on a non-BPS domain wall
with many tachyon fields living in its worldvolume. We found that in the thin wall limit the action is equivalent to the action of N non-BPS
parallel D3-branes in a flat five-dimensional bulk. The tachyon potentials can be sufficiently flat for large tachyon masses. One of the main
attempting considered here is to look for suitable inflaton potentials in superstring settings by searching for tachyon potentials in the non-BPS 
sector of a supergravity theory that is known to be the low energy limit of superstrings. Other attempting considering brane inflation have also
been considered \cite{dvali_tye,17,18}. We considered a supergravity 
theory with a scalar potential developing a $Z_2\times Z_2$ symmetry. This produces a non-BPS domain wall that can be interpreted as 
N non-coincident non-BPS D3-branes. As a future perspective one could also consider scalar potentials with, for instance, $Z_3$ symmetry in
the supergravity bosonic sector to account for tachyon fields in intersecting D3-branes system, where tachyon fields may interact and inflaton potentials even more realistic may appear.

{\bf Acknowledgments.} We would like to thank CNPq, CAPES, and PNPD/PROCAD -
CAPES for partial financial support.

\end{document}